\documentstyle[twocolumn,epsf,aps,prb,floats]{revtex}
\begin{document}
\draft

\twocolumn[\hsize\textwidth\columnwidth\hsize\csname @twocolumnfalse\endcsname

\title{Stripe stability in the extended $t$-$J$ model on planes
and four-leg ladders}

\author{T. Tohyama$^1$,  C. Gazza$^2$,
C. T. Shih$^3$, Y. C. Chen$^4$, T. K. Lee$^5$, S. Maekawa$^1$,\\
and E. Dagotto$^2$}

\address{$^1$ Inst. for Materials Research, Tohoku Univ., Sendai
980-8577, Japan}
\address{$^2$ National High Magnetic Field Lab and Dept. of Physics,
Florida State Univ., Tallahassee,\\
FL 32306, USA}
\address{$^3$ National Center for High-Performance Computing, Hsinchu,
Taiwan}
\address{$^4$ Dept. of Physics, Tunghai Univ., Taichung, Taiwan}
\address{$^5$ Inst. of Physics, Academia Sinica, Nankang, Taipei, Taiwan}

%\date{\today}
\date{Received 30 September 1998}

\maketitle

\begin{abstract}

The tendencies to phase-separation and stripe formation of the
$t$-$J$ model on planes and four-leg ladders have been here
reexamined including hole hopping terms $t'$, $t''$ beyond
nearest-neighbor sites.  The motivation for this study is
the growing evidence that such terms are needed for a
quantitative description of the cuprates. Using a variety of
computational techniques it is concluded that the stripe
tendencies considerably weaken when experimentally realistic
$t'<0$, $t''>0$ for hole-doped cuprates are considered.
However, a small $t'>0$ actually enhances the stripe formation.

\end{abstract}
\pacs{PACS numbers: 74.20.Mn, 74.25.Dw}
%\vskip2pc
]
\narrowtext

Growing experimental evidence suggests the existence of static
stripe order in a variety of transition metal oxides,
including hole-doped $\rm La_2 Ni O_4$ and
$\rm La_{1.6-x} Nd_{0.4} Sr_x Cu O_4$.\cite{stripes1}
In other cuprates, such as $\rm La_{1.85} Sr_{0.15} Cu O_4$
and $\rm Y Ba_2 Cu_3 O_{6.6}$, the magnetic scattering is
consistent with the presence of dynamic antiphase
antiferromagnetic (AF) domains.\cite{tranquada}
The origin of these charge inhomogeneities is controversial.
Some authors believe they are caused by Jahn-Teller (JT)
distortions.\cite{JT}  Others favor a purely electronic
explanation.  For instance, hole domain walls were observed
in Hubbard model Hartree-Fock calculations.\cite{zaanen}
In addition, computational calculations for the two-dimensional
(2D) $t$-$J$ model have found striped tendencies consistent
with neutron scattering experiments.\cite{tj,white2}
The existence of diagonal domain walls in four-leg
$t$-$J$ ladders has also been reported,\cite{white} adding
to the expected strong similarities between ladders and
planes.\cite{science}  A third possibility is based on
frustrated phase separation (PS) where the stripes arise
from a combination of a short-range attraction and
long-range Coulomb repulsion.\cite{emery}

In parallel to these developments recent angle-resolved
photoemission (ARPES) experiments addressed the one-particle
spectral function of the {\it undoped} insulator
$\rm Sr_2 Cu O_2 Cl_2$.\cite{wells}  The overall bandwidth
and features along the $(0,0)$-$(\pi,\pi)$ direction were
found in agreement with theoretical $t$-$J$ model
predictions.  However, the results along $(0,\pi)$-$(\pi,0)$
were puzzling since the ARPES quasiparticle-like peak has a
clear energy maximum at $(\pi/2,\pi/2)$, while in the $t$-$J$
model this line is almost flat.  This difference is important
and needs to be addressed.

The main explanation proposed for the $t$-$J$-ARPES discrepancy
is based on the relevance of corrections in the form of
electronic hopping amplitudes beyond the nearest-neighbor (NN)
contribution.  The importance of these terms was recognized
from the analysis of the electronic structures of
cuprates.\cite{hybertsen}  In all these calculations it was
concluded that for a proper description of cuprates a next-NN
(NNN) hopping of strength $t'$ along the plaquette diagonal
was necessary.\cite{laughlin} For hole-doped cuprates $t'$ has
been systematically found to be of $negative$ sign in contrast
to the NN-hopping amplitude $t$ with positive sign, and of
about 20\% to 40\% its magnitude.  Electron-doped cuprates
need $t'>0$.\cite{tohyama}  Note that in regions where AF
correlations are important, $t$ is renormalized to smaller
values while $t'$ is not severely affected, thus enhancing the
relevance of such NNN corrections.  In addition, soon after
the $\rm Sr_2 Cu O_2 Cl_2$ ARPES data\cite{wells} became
available it was reported by Nazarenko {\it et al}.\cite{nazarenko}
that including a $t'<0$ NNN-hopping the agreement
theory-experiment was noticeably improved at all momenta.
A similar result was also found by Lee and Shih.~\cite{lee}
Further work confirmed and improved this initial approach,
showing that with the addition of an extra NNN $t''$--term
connecting sites at distance $2a$ ($a$= lattice spacing) the
results improved even more.\cite{extra}  Actually with these
NNN corrections the hole-density dependence of the $t$-$J$
spectra of small clusters was found in agreement with ARPES
data.\cite{kim}  In Ref.~14 it was remarked that there is no
symmetry argument favoring the special case $t'=0$, as it
occurs in gauge theories where local symmetries and
renormalizability arguments fix the Hamiltonian.  Since the
existence of corrections to the $t$-$J$ model are natural, the
assumption  $t' = 0$ is mainly aesthetical.

The goal of the present paper is to address the much discussed
tendency of the $t$-$J$ model to phase-separate and/or form
stripes performing the calculations in the presence of a
realistic nonzero NNN-hopping amplitude.
The $t$-$J$ Hamiltonian employed here is defined as
$$
H_{tJ} = -t \sum_{{\langle {\bf ij} \rangle}\sigma} (c^\dagger_{{\bf i}
\sigma} c_{{\bf j}\sigma} + {\rm H.c.}) + J \sum_{\langle {\bf ij} \rangle}
( {{{\bf S}_{\bf i}}\cdot{{\bf S}_{\bf j}}} - n_{\bf i} n_{\bf j}/4 ),
%% \eqno(1)
$$
where $\langle {\bf ij} \rangle$ are NN sites. No doubly
occupancy is allowed, $t$ is defined as positive, and the rest
of the notation is standard.  The contribution of the NNN terms is
$$
H_{t't''} = -t' \sum_{{\langle\langle {\bf km} \rangle
\rangle}\sigma} c^\dagger_{{\bf k}
\sigma} c_{{\bf m}\sigma}  -t'' \sum_{{\langle\langle\langle {\bf nr}
\rangle\rangle\rangle}\sigma} c^\dagger_{{\bf n}
\sigma} c_{{\bf r}\sigma} + {\rm H.c.},
%% \eqno(2)
$$
where $\langle \langle {\bf km} \rangle \rangle$ denote a pair
of sites along the diagonals of the elementary plaquettes, and
$\langle \langle\langle {\bf nr} \rangle \rangle \rangle$ are
pairs of sites located at distance $2a$ along the main axis.

\begin{figure}[t]
\epsfxsize=8.0cm
\centerline{\epsffile{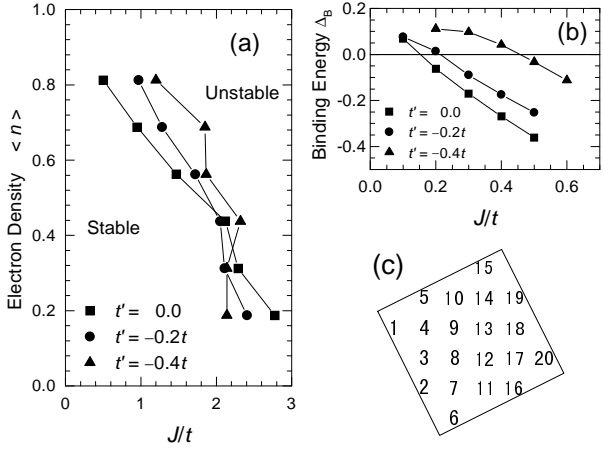}}
\vspace{2mm}
\caption{(a) Stable and unstable regions of the extended $t$-$J$
model on the $\langle n \rangle$-$J/t$ plane.  The results are
obtained exactly on a $4 \times 4$ cluster with ``averaged''
boundary conditions (see text); (b) binding energy vs $J/t$ of the
extended $t$-$J$ model on a $4 \times 4$ cluster, averaged boundary
conditions, and using the NNN hoppings indicated;
(c) site labeling used for the 20-site cluster.}
\end{figure}

To find information about the influence of $t'$ and $t''$ on PS
and stripe formation on the $t$-$J$ model, here a variety of
computational techniques have been employed.  Let us start the
analysis using the exact diagonalization (ED) method on planar
systems.  On 2D $t$-$J$ clusters evidence in favor\cite{hellberg2}
and against\cite{shih} PS at small $J/t$ has been recently
presented.  Our goal is not to add to this discussion, but rather
to follow some of the approaches proposed in those papers and
find the influence of NNN terms on the results.  Let us start
using ground state (GS) energies calculated by averaging with
equal weight over a large number of twisted boundary conditions
to reduce size effects.\cite{poilblanc}  Using this method
the GS energies corresponding to the electronic density
$\langle n \rangle =1$, the density under investigation
(with $N_e$ electrons), and the density corresponding to
$N_e+2$ electrons were obtained for increasing values of $J/t$.
When the three energies lie on a single line, PS occurs between
half-filling and the average between the densities $N_e/N$ and
$(N_e+2)/N$.\cite{hellberg}

Figure~1(a) illustrates the influence of a NNN hopping on the PS
tendencies of the $t$-$J$ model using a $4 \times 4$ cluster.
The largest density shown here corresponds to 13/16 and was
obtained with information from 12, 14, and 16 electrons.
For $t'=0$ the line separating the stable and unstable regions
converge to a very small $J/t$ as the density reaches
1.0.\cite{hellberg} However, even for an apparent ``small''
NNN hopping such  as $t'=-0.2t$, the PS line now converges
towards a larger $J/t$ at half-filling. The effect is similar
for $t'=-0.4t$.  Although these small-cluster results should
be considered only qualitative, the tendencies observed are
clear and in agreement with a variety of calculations reported
below.  Then, it is apparent that a nonzero $t'<0$ amplitude
reduces the tendencies towards PS in the $t$-$J$ model near
half-filling and small $J/t$.

Figure~1(b) contains the $\langle n \rangle = 1$ binding energy
defined as $\Delta_B=E(2)+E(0)-2E(1)$, where $E(n)$ is the GS
energy (averaged over boundary conditions) for $n$ holes.
The results show that a $t'<0$ term reduces the attraction
between holes. This is correlated with a reduction of the
probability of stripe formation and PS [Fig.~1(a)] caused by
the short-range AF-induced attraction.  The pairing region
is pushed up in couplings by a $t' < 0$ since it follows
PS.\cite{dago}  Results qualitatively similar to those shown
in Figs.~1(a) and 1(b) have been obtained using 18 and 20-sites
clusters and other values of $t'$.\cite{strong}

\begin{figure}[t]
\epsfxsize=8.0cm
\centerline{\epsffile{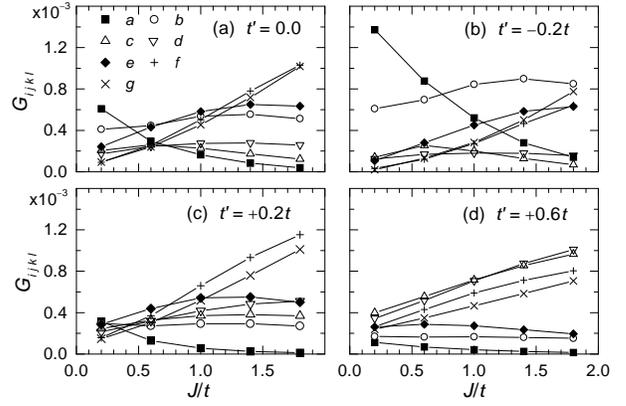}}
\vspace{2mm}
\caption{$G_{ijkl}$ for the seven configurations with the largest
weight in the ground state (see text) vs $J/t$. (a) Results for
$t'=0.0$; (b) same as (a) but for $t'=-0.2t$; (c) same as (a) but
for $t'=+0.2t$; (d) same as (a) but for $t'=+0.6t$.}
\end{figure}

To search for the dominant GS-hole configurations the matrix
element
$G_{ijkl} =$ $ \langle 0 | n_h(i)n_h(j) n_h(k) n_h(l) | 0 \rangle$
was calculated.\cite{tj} Here $(i,j,k,l)$ denote four sites of the
20-site cluster following the labeling convention shown in
Fig.~1(c).  The product of the four hole-number operators $n_h$
is a projector that provides the GS weight of the configurations
with the holes at $(i,j,k,l)$. The configurations with the largest
$G_{ijkl}$ are
$a=(4,7,14,17)$ (unbounded holes),
$b=(3,7,13,17)$ (pairs of diagonally bounded holes),
$c=(6,7,13,14)$ (NN-hole pairs),
$d=(6,7,13,18)$ (another NN-hole pair configuration),
$e=(2,8,13,19)$ [four-hole stripe along the (1,1) direction],
$f=(1,2,3,6)$ (another type of four-hole domain hole), and
$g=(3,8,12,17)$ [four-hole stripe along the (1,0) direction].
Figure~2(a) contains the results in the absence of NNN hoppings.
In this case three regimes can be identified: (1) at small $J/t$
the holes are unbounded; (2) for $J/t \sim 0.5$ holes form pairs;
and (3) at $J/t$ above 0.7 ``diagonal'' stripes are preferred,
as observed first in Ref.~5. Although these results are
qualitative, the tendencies towards (1,1)-stripe formation are
clear and also in agreement with four-leg ladder
calculations.\cite{white2}

Figure~2(b) contains the results found for the same hole
configurations but now using $t' = -0.2t$. Once again, in spite
of the naively ``small'' value of $t'$ its influence on GS
properties is important.  The $(1,1)$ stripes are no longer
competing with hole pairs and unbounded holes.  Now the most
relevant stripe configuration is the (1,0) stripe which dominates
only for $J/t \sim 1.9$ or larger.  Results similar to those
shown in Fig.~2(b) have been obtained using a variety of
$t'<0$ and $t''>0$ amplitudes.  Thus, it is clear that the
stability of the stripes is sensitive to the presence of
NNN-hopping amplitudes.  Since such hoppings are expected
to be realistic, the presence of stripes in electronic models
for the cuprates with short-range interactions is called into
question.

It is interesting to note that using $t' > 0$, i.e., the ``wrong''
sign for hole-doped cuprates but relevant for electron-doped
cuprates,\cite{tohyama} the tendencies to stripe formation are
actually {\it enhanced} roughly in the small window
$0 < t'/t < 0.2$ at all values of $J/t$.  Now the crossing point
between configurations $a$ and $f$ appears in Fig.~2(c) at
$J/t \sim 0.4$, while in Fig.~2(a) it occurred at $J/t \sim 0.7$.
Then, a small and positive $t'/t$ can be used as a test ground
of electronic models with tendencies towards stripe formation.
As $t'/t$ grows further stripes become unstable again, and
actually for $t'/t$ around 0.5 or larger the configurations
with NN-hole pairs dominate for all the values of $J/t$
explored here [Fig.~2(d)].

To understand the different influence of the sign of $t'$ on
the $t$-$J$ model phase diagram, a discussion on the subtleties
involving bare versus renormalized parameters is needed.
It is known that  at $t'=0$, the effective NN-hopping amplitude
is dramatically reduced at half-filling since intersublattice
hole hopping distorts the AF background. In this same regime
effective nonzero $t'$ and $t''$ amplitudes are generated,
as deduced from the one-hole dispersion.\cite{massimo}  The sign
of this effective $t'$ which gives mobility to the dressed hole
is $negative$, a well-known fact which manifest itself in the
minimum of the hole-quasiparticle band at $(\pi/2,\pi/2)$.
Adding a bare $t'$ term of the same sign will enhance the hole
mobility substantially, to the point where hole superstructures
become unstable. This amplification of effects explains the
results of Figs.~1(a) and 2(a) and 2(b).  However, the addition of a bare
$t'$ in the Hamiltonian of {\it positive} sign can lead to a
cancellation of effects, and a concomitant reduction of the hole
mobility near half-filling. This will indirectly favor hole
clustering since such structures arise from a competition of
the potential energy gained by AF attraction and the kinetic
energy.  Spin correlations are enhanced with respect to
$t'=0$,\cite{tohyama} since poorly mobile holes cannot scramble
them.  When $t'/t$ is increased further eventually holes should
become mobile again and the stripe tendencies will diminish.
Figure~2(d) shows $G_{ijkl}$ now for $t'/t=+0.6$. The dominant
configurations have NN-hole pairs instead of stripes at all
values of $J/t$ shown.

\begin{figure}[t]
\epsfxsize=8.0cm
\centerline{\epsffile{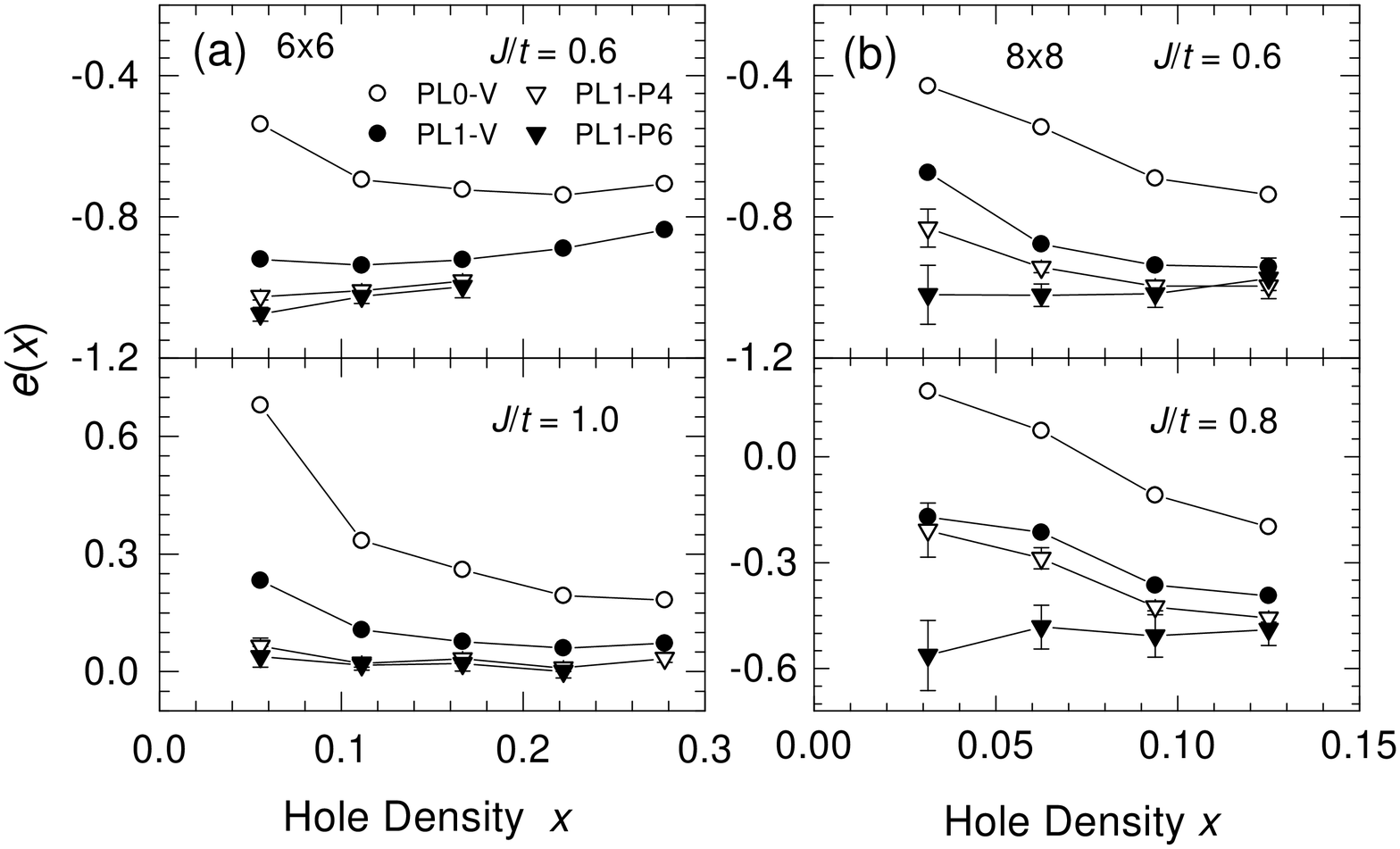}}
\vspace{2mm}
\caption{(a) $e(x)$ (see text) vs $x$ for several powers in the PL
method, two values of $J/t$, and  using a $6 \times 6$ cluster
with $t' = -0.2t$; (b) same as (a) but using a $8 \times 8$ cluster.}
\end{figure}

The GS energy of the $t$-$J$ model supplemented by a NNN term
with $t'=-0.2t$ has also been calculated using the Power
Lanczos (PL) method.\cite{chen}  To search for indications of
PS, the approach used in Refs.~19 and 26 is applied, namely
the energy of the PS state is written as $E=N_s e_H + N_h e(x)$,
where $N_s$ is the number of sites, $N_h$ the number of holes,
$e_H$ the energy per site of the Heisenberg model, $x$ the hole
density in a hole-rich phase, and $e(x) = [ e_h(x)-e_H ]/x$.
If at a fixed coupling
$J/t$,  $e(x)$ is found to have a minimum at some density $x_m$
and the overall density is smaller than $x_m$,
then the system phase
separates between a hole-free phase and a hole-rich one with
density $x_m$.  In Fig.~3 $e(x)$ is plotted vs $x$ using
$6 \times 6$ and $8 \times 8$ clusters with periodic boundary
conditions (open-shell configurations).  The energies denoted by
PL0-V correspond to results using an optimized trial wave
function taken from the set of Gutzwiller and resonant valence
bond (RVB) wave functions.\cite{shih,chen}  PL1-V denote
improved results now using the first Lanczos step applied to the
previously optimized wave function.\cite{adriana}  PL1-Pn
correspond to further improvements resulting from the
application of $n$ powers of the Hamiltonian over the
PL1-V wave function.  For additional details the reader is
referred to Refs.~19 and 25.

At $J/t=1.0$ and after the application of six powers,
the minimum of $e(x)$ in Fig.~3(a) is found to be at
$x = 0.22$.  At $J/t=0.6$, the minimum shifts with increasing
powers towards the smallest density studied here namely
$x = 0.056$.  Similar results were obtained for the
intermediate coupling $J/t=0.8$ (not shown).  In addition,
using an $8 \times 8$ cluster analogous trends are observed
as shown in Fig.~3(b), although with larger error bars due to
the sign problem.  The minimal values of $e(x)$ are at the
lowest doping density for $J/t=0.6$ and 0.8.  Therefore it is
concluded that the critical $J_c/t$ for PS in the low hole
density limit is at least $\approx0.8$, which is larger than
in the $t'=0$ case where $J_c/t\approx 0.6$.\cite{shih}
The trends found here are qualitatively the same as observed
using ED [Fig.~1(a)], namely a $t'<0$ moves the PS region
towards larger $J/t$'s.

\begin{figure}[t]
\epsfxsize=7.0cm
\centerline{\epsffile{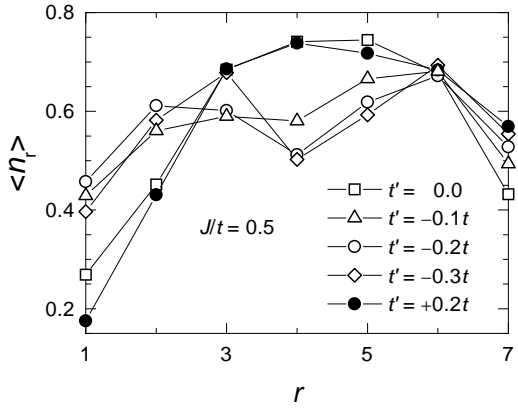}}
\vspace{2mm}
\caption{Rung-density $\langle n_r \rangle$ vs $r$ on a
$4 \times 14 $ cluster, $J/t=0.5$, 8 holes, studied with DMRG and
open boundary conditions.  Shown are only half the rungs, the rest
is found by reflection. $r=1$ ($r=7$) is the end (middle) of
the cluster.  Results for several $t'$'s are shown.}
\end{figure}

Since the most robust computational evidence of stripe
formation in one-band electronic models actually comes from
the density-matrix numerical renormalization group (DMRG)
studies of four-leg $t$-$J$ ladders, let us complete our
analysis by studying the same clusters as in Ref.~7, following
a similar methodology, but now adding NNN terms.  The DMRG
results reported here were obtained using $J/t=0.5$, $m=500$
states, and a truncation error $\sim 1 \times 10^{-4}$.
In Fig.~4 the rung density $\langle n_r \rangle$ (i.e., the sum
of the four-site densities forming a rung) is shown using a
$4 \times 14$ cluster and 8 holes. Results for just half the
lattice are provided for simplicity, since the rest are
obtained by reflection.  At $t'=0$, previous
results\cite{white} were reproduced as a test. They present a
single broad peak indicative of GS clustering tendencies.
Analyzing the hole-hole correlations the $(1,1)$ stripes were
found to have a large weight in the GS.\cite{white}  However,
when the NNN amplitude is turned on, the effect is weaken.
Consider, for example, $t'=-0.2t$: now a two-peak structure is
observed which is more suggestive of hole pairing than of stripe
formation (note there are four holes in average on the portion
of the cluster shown in Fig.~4).  The effect is further
enhanced for $t' = -0.3t$ where the two peaks are sharper.
Here the holes were found to be unbounded residing in pairs
at the extremes of the same rung. Then, as $|t'|$ grows a
rapid transition from stripes to unbounded holes is observed.
The melting of the stripes (assumed to be signaled by the
melting of the single-peak in $\langle n_r \rangle$) roughly
occurs at $t'\sim -0.1t$.  More realistic values of the
coupling are difficult to study as accurately as at $J/t=0.5$.
However, it is expected that the short-range AF attraction
will become weaker as $J/t$ is reduced, and the tendencies
to stripe formation will also be weaker in this regime.
Note that once again a $t'>0$ maintains the stripe structure
(single peak shown in Fig.~4 for $t'=+0.2t$), as in planar systems.

Summarizing, using a variety of computational techniques it has been
shown that the tendencies to phase separation and stripe formation
previously reported on planes and four-leg ladders are actually
substantially weaken once realistic NNN-hopping amplitudes are
added to the $t$-$J$ model.  The reason is that $t'<0$ terms
enhance appreciably the mobility of holes, melting hole
superstructures.  Reciprocally, using a (small) $t'>0$ the stripes
become more stable providing an interesting model for the analysis
of their properties.

C.G. thanks the support of CONICET (Argentina).
T.K.L. was supported by the NSC grant No. 87-2112-M-001-016.
E.D. was supported by the NSF grant No. DMR-98-14350.
S.M. was supported by the CREST and NEDO.
\medskip

%\vfil


\begin{references}
\bibitem{strong}On the $4\times 4$ cluster calculations with
$t'' \neq 0$ have strong size effects, and they were not pursued here.
\end{references}
\end{document}